\documentclass[onecolumn, aps, prd, groupadress, nofootinbib, superscriptaddress, showkeys, showpacs]{revtex4-1}

\usepackage{amsmath, amsfonts, amssymb, mathrsfs}
\usepackage{mathtools,nccmath}
\usepackage[geometry]{ifsym}

\usepackage{scalerel,graphicx}
\usepackage[font=small,labelfont=bf]{caption}
\usepackage[font=small,labelfont=bf,textfont=normal,justification=justified,singlelinecheck=false]{subcaption}

\usepackage[T1]{fontenc} 

\usepackage{cmbright}

\usepackage{latexsym}

\usepackage[hyperindex]{hyperref}
\hypersetup{breaklinks=true, hyperfootnotes=true, pagecolor=white, colorlinks=true}
\usepackage[all]{hypcap}

\usepackage{tensor}

\newcommand{\Fkl}{\tensor{F}{_\kappa_\lambda}}

\newcommand{\phimn}{\tensor{\Phi}{_\mu_\nu}}

\begin{document}

\title{A new bifurcation in the Universe}

\

\author {A.~E.~S. Hartmann and M. Novello}
 \affiliation{
Centro de Estudos Avan\c{c}ados de Cosmologia / CBPF \\
Rua Dr. Xavier Sigaud 150, Urca 22290-180 Rio de Janeiro, RJ-Brasil}
\date{\today}

\vspace{2 cm}

\date{ \today}

\begin{abstract}
We show that the combined system of general relativity with a non minimally coupled electromagnetic field presents a bifurcation in a cosmical framework driven by a cosmological constant. In the same framework we show the existence of states such that the resulting combined energy (the sum of the minimally and the non minimally coupled energy momentum tensor of the electromagnetic field) vanishes in a sort of violation of the action-reaction principle.\\[1ex]
\end{abstract}

\maketitle

\section{Introduction}
In the early 1984, M. Novello and Ligia Rodrigues \cite{ligianovello} analysed the presence of a bifurcation in the universe driven by a viscous fluid. In this kind of classical description, the isotropic pressure of the stress-energy tensor is written as a polynomial of the expansion factor $\theta$,
\begin{align}
\widetilde{p}\, = \, p\, + \,\alpha\,\theta + \,\beta\, \theta^2.
\end{align}
As they showed, dissipative processes can provoke the appearance of bifurcation in autonomous non-linear differential equations, as is the case of the equations of general relativity in a spatially isotropic and homogeneous universe. 
The present paper constitutes the return to the open questions of this kind of indeterminacy in cosmological solutions in a very distinct context.

In the standard cosmological description of the electromagnetic fluid as the main source of primordial gravitational field, a sort of average procedure must be made once the geometry is identified to a spatially homogeneous and isotropic metric. This led to describe the electromagnetic effect as nothing but a cosmical fluid with density of energy $\rho = 1/2 \, (E^{2} + B^{2})$ and pressure $ p = 1/3 \, \rho$ given by the traceless property.

The exam of the global properties of the space-time suggests the question: 
What is the effect on this description if the coupling of the electromagnetic field to gravity is non minimal and contains functions of the curvature? We shall see that  with the same sort of average procedure as in the standard minimal coupling, there is a sea of bifurcations when general relativity is combined with non-minimal interaction of gravity with linear electromagnetic field.

\section{Duality rotation in minimal and nonminimal couplings}

Consider the Einstein and Maxwell theories described by the minimally coupled Lagrangian

\begin{align}
L = \frac{1}{2 \,k} \, R - \frac{c^{2}}{4} \, F, \label{MC}
\end{align}
in which $F := F_{\mu\nu}\, F^{\mu\nu}=2({B}^2-{E}^2)$. The field equations for the metric are given by  (see the Appendix I for definitions and conventions)
\begin{align}
R_{\mu\nu} -\,\frac{1}{2}\, R\, g_{\mu\nu} ~= -\kappa T_{\mu\nu}^{\gamma},
\end{align}
where the curvature tensor is defined as
\begin{align}
v_{\alpha;\mu;\nu} - v_{\alpha;\nu;\mu} ~=~ R^{\alpha}{}_{\beta\mu\nu}v^\beta
\end{align}
and the Maxwell energy-momentum tensor is
\begin{align}
T_{\mu\nu}^{\gamma} ~=~ \phimn  + \frac{1}{4} \, F \, g_{\mu\nu},
\end{align}
with $\phimn:= F_{\mu\alpha} \, F^{\alpha}{}_{\nu}$.

It is well known that the Maxwell equations are invariant under the global duality rotation $\mathscr{R}$($\alpha$) of the electromagnetic field, stated by the relations

\begin{align}
\begin{split}
F_{\mu\nu} \quad \overset{\mathscr{R}(\alpha)}{\longmapsto}\quad F^{'}_{\mu\nu} ~&= \,\cos\alpha\, F_{\mu\nu} + \,\sin\alpha\, \overset{*}F{}_{\mu\nu} \\[1ex]	
\overset{*}F{}_{\mu\nu} \quad \overset{\mathscr{R}(\alpha)}{\longmapsto} \quad\overset{*}F{}^{'}_{\mu\nu} ~&= - \,\sin\alpha \, F_{\mu\nu} +\cos\alpha \, \overset{*}F{}_{\mu\nu},
\end{split}
\end{align}

with

\begin{align}
 \overset{*}F{}^{\mu\nu} :=\frac{1}{2}\tensor{\eta}{^\mu^\nu^\kappa^\lambda}\Fkl.
\end{align}
Moreover the Lagrangian (\ref{MC}) is invariant in first order on the dual angle,

\begin{align}
\begin{split}
F^{'}_{\mu\nu} ~&\approx~  F_{\mu\nu} + \alpha \, \overset{*}F{}_{\mu\nu} \\[1ex]
\overset{*}F{}^{'}_{\mu\nu} ~&\approx~ - \alpha \, F_{\mu\nu} + \, \overset{*}F{}_{\mu\nu},
\end{split}
\end{align}

from which it follows, up to a total divergence

\begin{align}
F' \approx F .
\end{align}

Furthermore there is no effect on the dynamics of the metric once the Maxwell's energy-momentum tensor $T_{\mu\nu}^{\gamma}$ is invariant under the global map $\mathscr{R}(\alpha)$,

\begin{align}
T_{\mu\nu}^{\gamma'} = T_{\mu\nu}^{\gamma}.
\end{align}
The proof of this is immediate and one should use the algebraic properties of the electromagnetic tensor shown in the Appendix I.

Now comes the question: is it possible to generalize the interaction between these two fields for a non-minimal coupling in which the curvature tensor appears in the interacting term with the electromagnetic field in such a way that the theory remains invariant under $\mathscr{R}(\alpha)$? The answer is yes and the corresponding interacting Lagrangian is provided by \cite{novello1987,novellobergliaffa}

\begin{align}
L_{int} = \zeta \, C_{\mu\nu} \, \Phi^{\mu\nu}, \label{NMC}
\end{align}

where $ \zeta$ is the nonminimal coupling constant with dimensionality of $ length^{2},$ and

\begin{align}
C_{\mu\nu} ~&:=~ R_{\mu\nu} - \frac{1}{4} \, R \, g_{\mu\nu}.
\end{align}

Note that this extra term $ L_{int}$ is invariant even in the case the rotating angle depends on spacetime. In other words, gravity is responsible for recovering the global invariance of electromagnetic field and introducing a new symmetry without counterpart in the Maxwell's theory. The Lagrangian of interaction (\ref{NMC}) can be written in the equivalent form
\begin{align}
C_{\mu\nu} \, \Phi^{\mu\nu} ~=~ R_{\mu\nu} \, T^{\mu\nu}_{\gamma}.
\end{align}

One can go a step further and instead of use the contracted Riemann tensor $ R_{\mu\nu}$ one can consider the conformal Weyl tensor $ W_{\alpha\beta\mu\nu} $ to describe nonminimal coupling. In this case there is a generalized dual rotation that acts both on the Weyl and Faraday tensors that leaves the interaction of the combined electromagnetic and gravitational fields invariant. It is shown in the appendix that such duality invariance occurs if the rotating angle for the spin-2 field, represented by the Weyl conformal tensor, is precisely twice the spin-1 rotating angle, that acts on the electromagnetic field. In the present paper we limit our analysis to the above $ L_{int}.$

Traditionally the discussion on the duality rotation of the electromagnetic field concerns to the exam of compatibility of magnetic monopoles with the quantum theory,  a question introduced by the seminal paper of Dirac in 1931 \cite{dirac1931}. Despite this, the present work follow another possible path.

\section{The field equations}

Let us start by choosing the action principle as given by

\begin{equation}
\delta\int \, \sqrt{-g} \,  \left(\frac{1}{2 \,k} \, R - \frac{c^{2}}{4} \, F + \frac{\zeta}{2} \, C_{\mu\nu} \, \Phi^{\mu\nu} + L_{m} \right) = 0
\label{14jan1}
\end{equation}
where $ L_{m}$ is the rest of matter contribution. The equation of the metric is given by

\begin{equation}
  R_{\mu\nu} - \frac{1}{2} \, R \, g_{\mu\nu} = - \, T_{\mu\nu}^{\gamma} - \zeta \,  Z_{\mu\nu} - T_{\mu\nu}^{m}
  \label{gr1}
  \end{equation}
where

\begin{align}
Z_{\mu\nu} ~&=~ - \, \frac{1}{2} \, g_{\mu\nu} \, ( R_{\rho\sigma} \, \Phi^{\rho\sigma} + \frac{1}{4} \, R \, F )  + R^{\varepsilon}{}_{(\mu} \, \Phi_{\nu)\varepsilon} + R^{\rho\sigma} \, F_{\rho\mu} \, F_{\nu\sigma} - \frac{1}{2} \, R \, \Phi_{\mu\nu} + \frac{1}{4} \, F \, R_{\mu\nu} \nonumber \\[1ex]
&\quad ~+ \frac{1}{2} \, \Phi^{\varepsilon}{}_{(\mu;\nu) ;\varepsilon}  + \frac{1}{4} \, F_{;\mu;\nu}   - \frac{1}{2} g_{\mu\nu}( \Phi^{\rho\sigma}{}_{;\sigma;\rho} + \frac{1}{2}\Box F ) - \frac{1}{2} \Box \Phi_{\mu\nu}.
\end{align}
The equation for the electromagnetic field is
\begin{align}
 F^{\mu\nu}{}_{; \nu} + \zeta \, ( C^{\mu}{}_{\epsilon} \, F^{\epsilon\nu} + F^{\mu}{}_{\epsilon} \, C^{\epsilon\nu} )_{; \nu} ~=~ 0.
 \end{align}

\section{The average procedure}

We are interested in analyze the effects of the non minimal coupling in the standard spatially isotropic and homogeneous metric. To be consistent with the symmetries of this choice of the metric,
an averaging procedure must be performed if electromagnetic fields
are to be taken as a source for the gravitational field, according to the standard procedure \cite{tolmanbook}. The definition of the average of a quantity $Q$ at a instant $t$ is given by
\begin{align}
    < Q > ~&=~ \lim_{V\rightarrow \infty} ~\frac{1}{V}\int \sqrt{-g}~ Q~d^3x,
    \label{Def:media}
\end{align}
with $V = \int\sqrt{-g} d^3x.$ As a consequence, the components of the electric $E_{i}$ and magnetic $B_{i}$ fields must satisfy the following relations:
\begin{align}
<{E}_i > ~=~ 0,\qquad
<{B}_i> ~&=~ 0,\qquad
<{E}_i\, {B}_j> ~=~ 0,
\\[1ex]
<{E}_i\,{E}_j> ~&=~ -\, \frac{1}{3} {E}^2
\,g_{ij},
\\[1ex]
<{B}_i\, {B}_j> ~&=~  -\, \frac{1}{3} {B}^2
\,g_{ij}.
%
\end{align}
where $E$ and $B$ depends only on time. Besides we assume that the time derivative operation commutes with the average procedure, that is we set

\begin{align}
\partial_{t} \, <{E}_i\,{E}_j> ~&=~  \, <\partial_{t}  \, (E_i\,E_j) >, \\[1ex]
\partial_{t} \, <{B}_i\,{B}_j> ~&=~  \, < \partial_{t}  \, (B_i\,B_j) >, \\[1ex]
\partial_{t} \, <{E}_i\,{B}_j> ~&=~  \, < \partial_{t}  \, (E_i\,B_j) >.
\end{align}

Using the above average values it follows that the Maxwell energy-momentum tensor $T_{\mu\nu}^{m}$
reduces to a perfect fluid configuration with energy density
$\rho_\gamma$ and pressure $p_\gamma$ given by
\begin{align}
<T_{\mu\nu}^{\gamma} > ~=~ (\rho_\gamma + p_\gamma)\,
v_{\mu}\, v_{\nu} - p_\gamma\, g_{\mu\nu}, \label{Pfluid}
\end{align}
where
\begin{align}
\label{RhoMaxwell} \rho_\gamma = 3p_\gamma = \frac{1}{2}\,(E^2 +
{B}^2),
\end{align}
Note that
\begin{align}
 < \Phi_{\mu\nu}> ~=~ \frac{2}{3} \, (\sigma^{2} + 1) \, X \, v_{\mu} \, v_{\nu} + \frac{1}{3} \, (\sigma^{2} - 2) \, X \, g_{\mu\nu},
\end{align}
where we have set
$ E^{2} = \sigma^{2} \, B^{2}.$    For simplicity, from now on we will write $ B^{2} = X.$

Let us look for the other part $  < Z_{\mu\nu} > .$

\subsection{$ Z_{\mu\nu}$ interpreted in terms of a perfect fluid}

In the case of non minimal coupling with gravity the extra term for the energy-momentum tensor $ Z_{\mu\nu}$ is rather involved once it contains terms depending on the curvature. Let us set for the metric the form

\begin{align}
ds^{2} = dt^{2} - a^{2}(t) \, (dx^{2} + dy^{2} + dz^{2})
\end{align}
In this case
\begin{align}
 R_{00} &= \dot{\theta} + \frac{1}{3} \, \theta^{2},\\
 R_{ij} &= \frac{1}{3} \,( \dot{\theta} +  \, \theta^{2}) \, g_{ij}, \\
R &= 2 \, \dot{\theta} +  \frac{4}{3}\, \, \theta^{2},
\end{align}
where we have defined the expansion factor $ \theta = 3 \, \dot{a}/a.$

Using these results one obtains

\begin{align}
< Z_{\mu\nu} > ~&=~ M \, g_{\mu\nu} + N \, \, v_{\mu} \, v_{\nu} + \frac{1}{3} \,(\sigma^{2} + 1) \, X_{, \epsilon ; (\mu} \,v_{\nu)} \, v^{\epsilon} \nonumber \\
&\quad + \frac{2}{9} \,(\sigma^{2} + 1) \, \theta\, X_{, ( \mu} \, v_{\nu)} + \frac{1}{3} \,(\sigma^{2} + 1)\, X \, R^{\alpha}_{( \mu} \, v_{\nu)} \, v_{\alpha} \nonumber \\
&\quad - \frac{1}{6} \,(\sigma^{2} + 1) \, X_{, \mu ; \nu} + \frac{1}{2} \,(\frac{\sigma^{2}}{3} - 1) \, X \, R_{\mu\nu}.
\end{align}
with $M$ and $N$ defined by
\begin{equation}
M ~=~ - \, \frac{1}{6} \, (\sigma^{2} + 1) \,\ddot{X} - \frac{5}{18} \, (\sigma^{2} + 1) \, \theta \, \dot{X} + \frac{1}{18} \, (1 - 5 \, \sigma^{2}) \, X \, \dot{\theta}
- \frac{2}{9} \, \sigma^{2} \, \theta^{2} \, X,
\end{equation}
and
\begin{equation}
N ~=~ - \, \frac{1}{3}  \,(\sigma^{2} + 1) \, \ddot{X} - \, \frac{1}{3} \, (\sigma^{2} + 1) \,\theta \, \dot{X} - \, \frac{2}{9} \,(\sigma^{2} + 1)\, X \, \dot{\theta} - \frac{2}{9} \, (\sigma^{2} + 1) \, X \, \theta^{2}.
\end{equation}

\subsection{The perfect fluid}

From what we have shown we can analyze the effects of the non minimal coupling of the electromagnetic field with gravity in a spatially homogeneous and isotropic geometry in terms of the standard equation of general relativity and a mixed fluid as in equation (\ref{gr1}). In order to compatibilize the theory with a spatially homogeneous and isotropic metric one must analyze its corresponding heat flux $ q_{\mu}$ and the anisotropic pressure $ \pi_{\mu\nu}$ which are defined for an arbitrary energy-momentum tensor $T_{\mu\nu}$ as
\begin{align}
 q_{\lambda} ~&=~ T_{\alpha\beta} \, v^{\beta} \, h^{\alpha}{}_{\lambda} \\[1ex]
\Pi_{\mu\nu} ~&=~ T^{\alpha\beta} \, h_{\alpha\mu} \, h_{\beta\nu} + p \, h_{\mu\nu}
\end{align}
where
\begin{align}
h_{\mu\nu} ~=~ g_{\mu\nu} - v_{\mu} \, v_{\nu}
\end{align}
and the pressure
\begin{align}
 p ~=~ - \, \frac{1}{3} \, T_{\mu\nu} \, h^{\mu\nu}.
 \end{align}

In the case of tensor $ Z_{\mu\nu}$ it is a rather long although direct calculation to show that both  $ q_{\mu}$ and $ \pi_{\mu\nu}$ vanish. This yields the remarkable consequence that we can write the coupled energy-momentum tensor $ Z_{\mu\nu}$ in the form of a perfect fluid
$$ Z_{\mu\nu} = (\rho_{z} + p_{z} ) \, v_{\mu} \, v_{\nu} - p_{z} \, g_{\mu\nu} $$
where
\begin{align}
\rho_{z}=  - \, \frac{1}{6} \, (\sigma^{2} + 1 ) \theta \, \dot{X} +  \frac{1}{3} \, \sigma^{2} \, \dot{\theta} \, X - \frac{1}{6} \,( \sigma^{2} + 1) \, \theta^{2} \, X,
\label{25121}
\end{align}
and
\begin{align}
p_{z}= \frac{1}{6} \, (\sigma^{2} + 1) \, \ddot{X} +  \frac{1}{3} \,( \sigma^{2} + 1) \, \theta  \, \dot{X} + \frac{1}{9} \, (2 \,\sigma^{2} + 1) \, \dot{\theta} \, X + \frac{1}{6} \,( \sigma^{2} + 1) \, \theta^{2} \, X .
\label{25122}
\end{align}

We will now show that this system admits a bifurcation point. Before this, let us make a short presentation of the mathematical scheme.

\section{Conditions for the bifurcation point}

Let $p$ and $q$ be two variables which characterize a physical
system - described in the phase plane $(p, q)$ - whose evolution in
time  $t$ is given by the planar autonomous system of differential
equations
\begin{align}
\begin{split}
\dot{p} &=f(p,q,\lambda ) \\
\dot{q} &=h(p,q,\lambda )\ , \label{DS}
\end{split}
\end{align}
in which $ f$ and $h$ are non-linear functions. The system is called autonomous because the right-hand side of (\ref{DS}) does not contain explicitly the time variable $t$. We have add to (\ref{DS}) a parameter $\lambda$  which has a given range ${\mathcal  M}$ on the real axis and distinguishes special interactions among parts of the physical system.

The states of equilibrium of the system (\ref{DS}) are given by the
points in the phase plane $(p, q)$  for which $f$ and $h$ are
simultaneously annihilated. A multiple equilibrium state is called a \textit{bifurcation point} if for a given value of the parameter, say $\lambda_{o},$  the topological behavior of the integral curves changes discontinuously
when passes the value $\lambda_{o}$.

This physical situation of instability of the singular point in the phase plane,
coupled to the aleatory character of the fluctuations which the
system can undergo, extinguishes almost completely the possibility of
predictions. In other words, the system arrived at the vicinity of a
bifurcation point evolves in a non-deterministic way, which is a
situation already implicitly contained in the equations used to
describe the system. The proof of this is based on Bendixson's theorem \cite{andronov} which
states that the Poincar\' e index - which is a kind of measure of the
topological properties in the neighborhood of the singular point -
of a multiple equilibrium state is given by the relation
\begin{align}
I_p =\frac{E-H}{2}+1 \ ,
\end{align}
in which $E$ and $H$ represent, in the phase plane, the number of
elliptical and hyperbolic sectors, respectively.

The sudden modification of the topological properties of the
integral curves of the system in the phase plane represents an abrupt change of behavior of the physical
system in the vicinity of the unstable point. The crucial
consequence of this is the appearance of non-deterministic features
in the metrical properties of the universe, even in a classical and non-statistical description.

\section{An example of cosmological solution with bifurcation}

Start by assuming that the expansion factor is a constant driven by $ L_{m}$ identified to a cosmological constant and the electromagnetic fluid is just a test field (in the next section we will show that there is a state such that the energy of the fluid interacting nonminimally with gravity, the $ Z_{\mu\nu} $ contribution is supressed). We look for the fundamental state of the cosmic fluid, that is we set
\begin{align}
 p +\rho = 0,
\end{align}
where
$$ \rho =\rho_{\gamma} + \zeta \, \rho_{z}$$
$$ p = p_{\gamma} + \zeta \, p_{z}.$$

which yields the differential equation
\begin{align}
\zeta \ddot{X} + \zeta\theta_0 \dot{X} + 4X ~=~0.
\end{align}

\begin{figure*}[!htb]
\centering
\caption{Dynamical system with $\theta_0>0$ (to left) and $\theta_0<0$ (to right).}
\begin{minipage}{.90\linewidth}
\centering
\subcaption{$\zeta>0$:}\vspace{-.1cm}
\includegraphics[width=.44\textwidth]{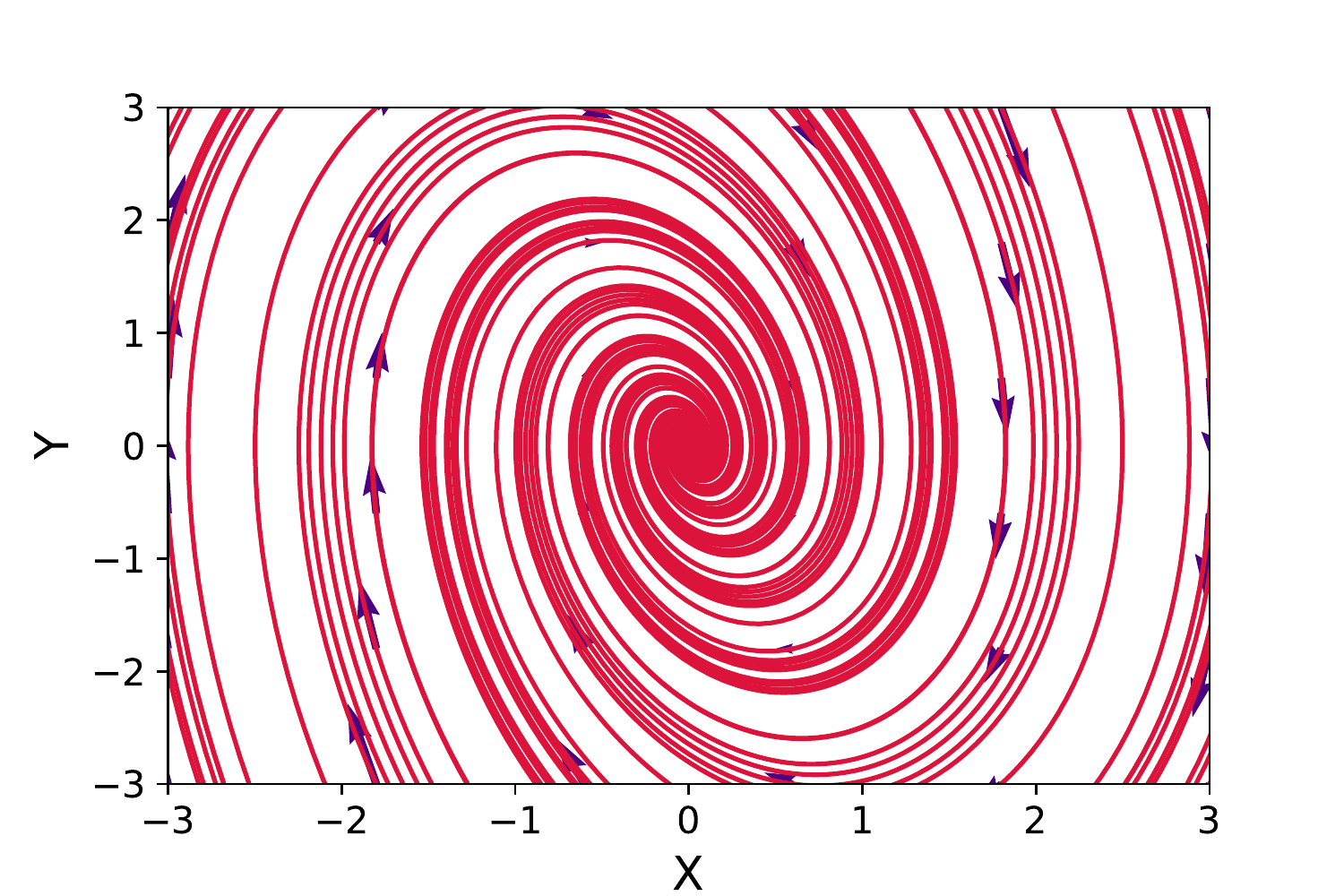}\hspace{.05cm}
\includegraphics[width=.44\textwidth]{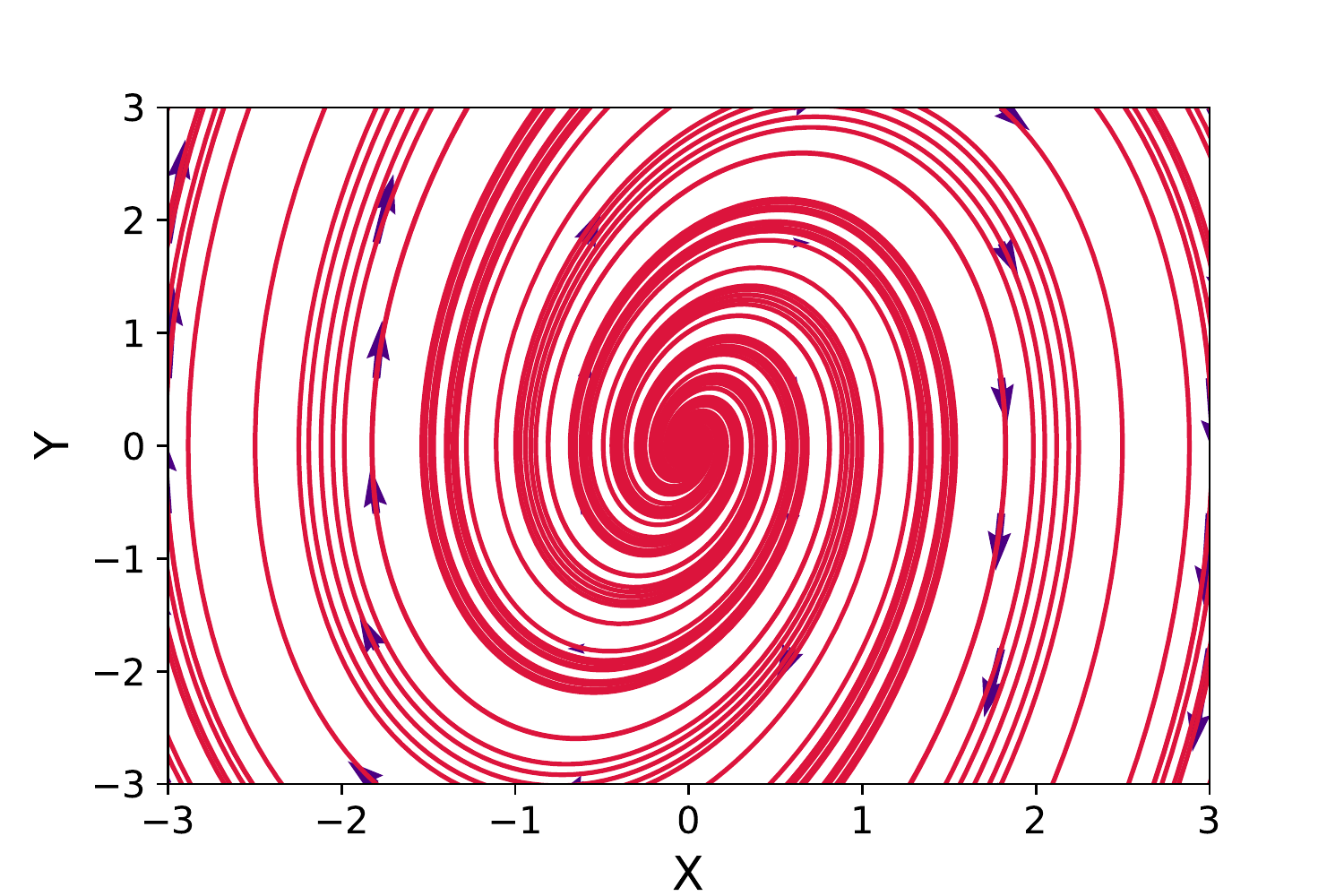}\vspace{.15cm}
\label{Fig.SD1a-b}
\end{minipage}
\begin{minipage}{.90\linewidth}
\centering
\subcaption{$\zeta<0$:}\vspace{-.1cm}
\includegraphics[width=.44\textwidth]{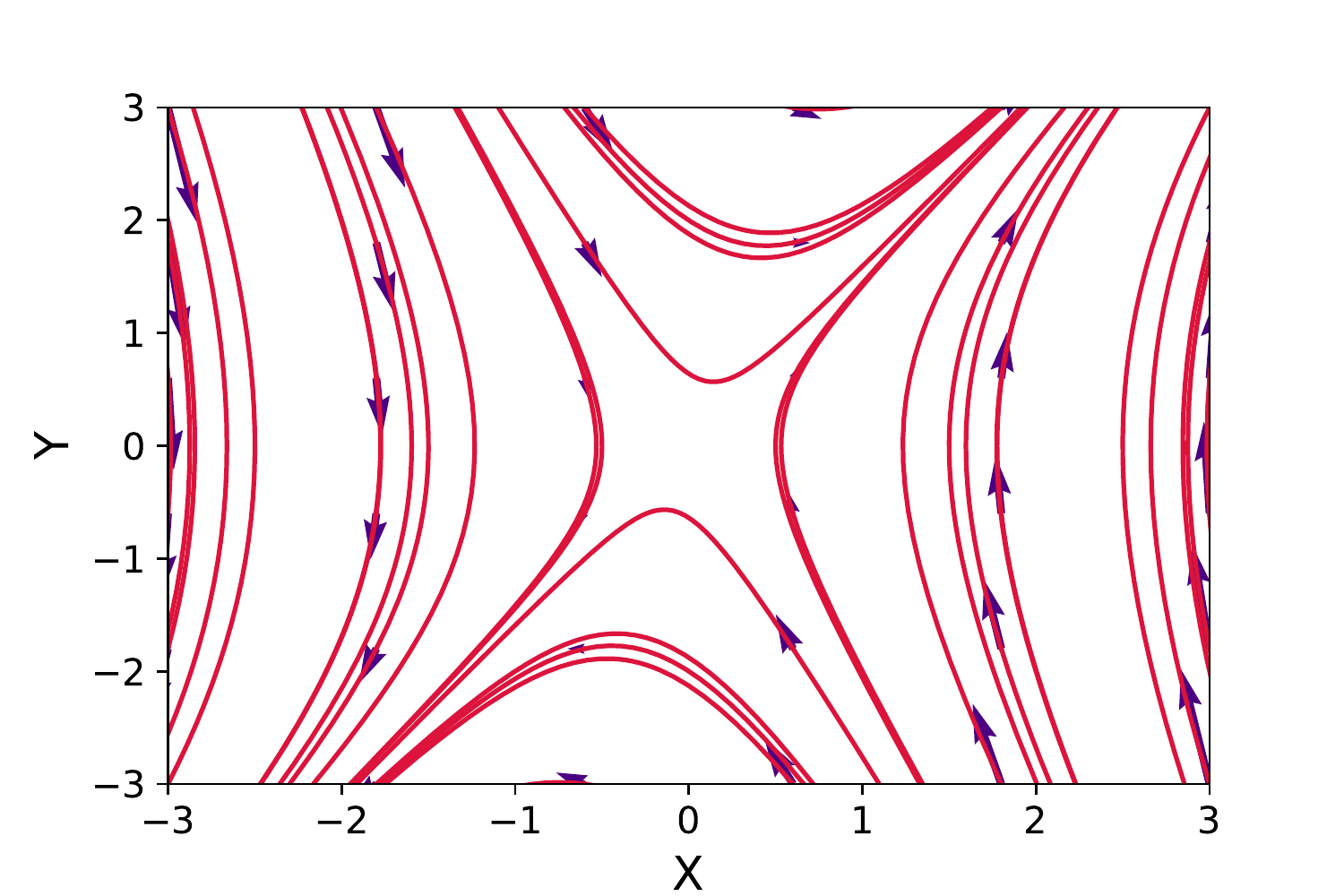}\hspace{.05cm}
\includegraphics[width=.44\textwidth]{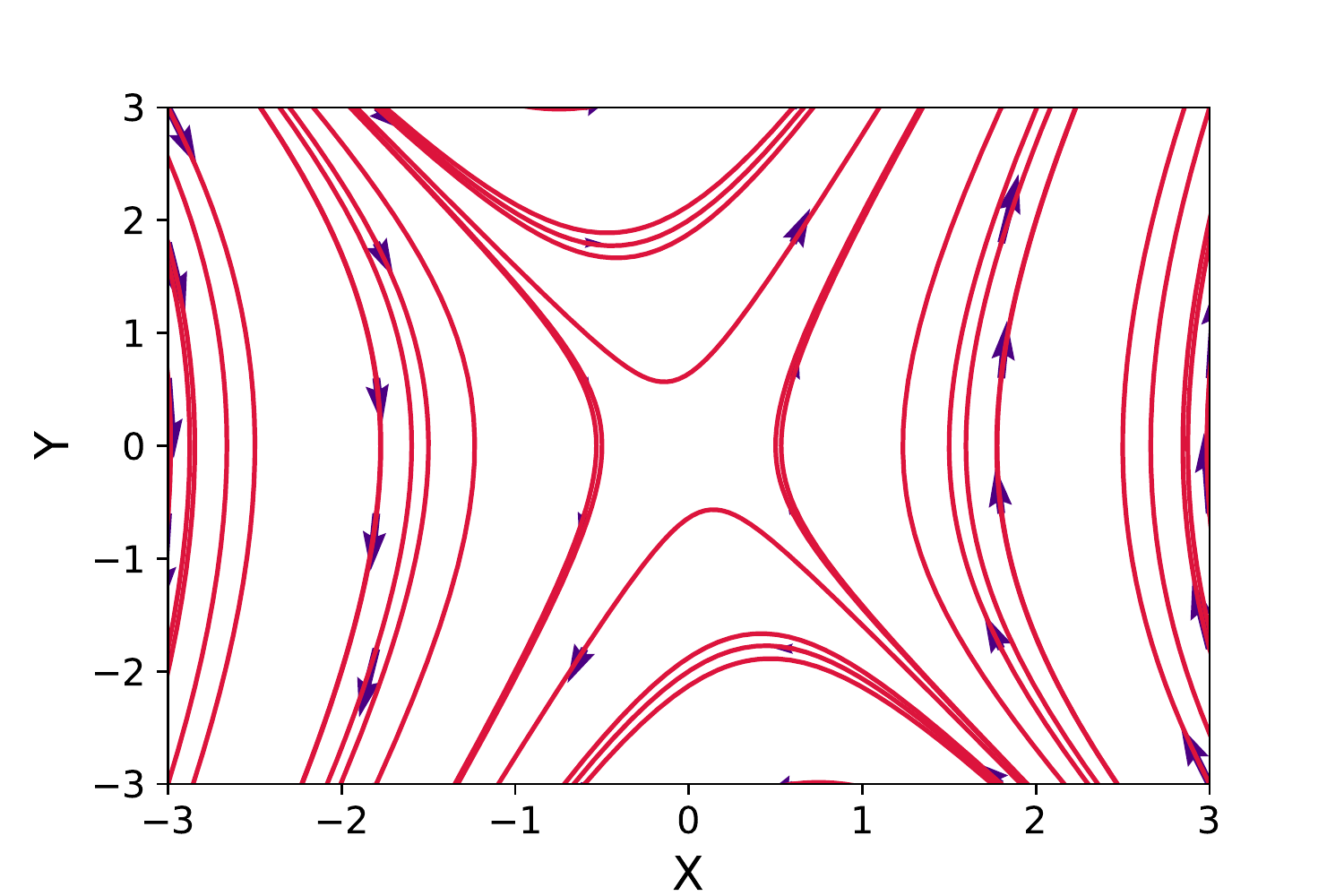}\vspace{.15cm}
\label{Fig.SD1c-d}
\end{minipage}
\end{figure*}

The analytical solution is immediate and no further investigation should be required from the mathematical viewpoint. However the richness of the cosmological questions leaves us for searching another property. In this vein we will investigate the global behaviour of such dynamics by asking: can the topology of the dynamical system describing such cosmological solution change in time? The most simple and direct qualitative analysis \cite{andronov}, \cite{novelloaraujo}  reveals the important properties of the planar autonomous system

\begin{align}
    \dot{X} ~&=~ Y \\
    \dot{Y} ~&=~ -\frac{4}{\zeta}X - \theta_0 Y.
\end{align}
There is a critical point at the origin $(X,Y)=(0,0)$. The corresponding matrix $\Delta = \partial (\dot{X},\dot{Y})/\partial(X,Y)$ is given by

\begin{align}
    \Delta ~=~ \left(\begin{array}{cc}
    0 & 1 \\
    -4/\zeta & -\theta_0
    \end{array}\right).
\end{align}
The determinant is positive when $\zeta>0$. In this case the equilibrium point is a focus, stable for $\theta_0$ positive and unstable for $\theta_0$ negative (as shown by the figures \ref{Fig.SD1a-b}). For $\zeta<0$ yields $det\Delta<0$ and consequently the equilibrium point is a saddle (see the Figures \ref{Fig.SD1c-d}). If $det\Delta\neq0$, the only one singular point is the origin \cite{andronov}.  The case

$$ \lim_{\zeta\longleftrightarrow\infty} \det \Delta ~=~ 0$$

implicates entire lines of singular points.

The index of Poincar\'e of a simple equilibrium state is $+1$ in the case of a node or focus and is $-1$ in the case of a saddle point. Thus this system contains potentially a change of topology that is the necessary condition for the existence of a bifurcation point \cite{andronov}.

Let us point out the dependence of the bifurcation point on the value of the non minimal coupling constant. Besides, this property of exhibiting bifurcation is very general, valid for almost all relationship between $ p $ and $ \rho$ except for the case of a "pure" electromagnetic radiation, that is,  $p= 1/3\rho$.
\section{Energy supressed by curvature or states that violate the action-reaction principle}

A very unexpected result of the property of reduction of the energy distribution of the non minimal coupling of the electromagnetic field to gravity to a perfect fluid as showed above concerns the possibility of a state such that the combined energy-momentum tensor of the free field plus the interacting term, interpreted as the equations of General Relativity (see equation (\ref{gr1})), does not drive the gravitational field. How is this possible? Let us show it.
From the average procedure the coupled system that remains to be solved reduces to the conservation of the energy $\rho_{m} + \rho_{z}$ and the evolution equation for the expansion factor. Let us analyze the case in which the gravitational field is due only to the extra matter represented by the term $ L_{m}$ in the total Lagrangian (\ref{14jan1}) by imposing the two conditions: i) the sum of the density of energy of the part of the electromagnetic field coupled minimally to gravity $\rho_{\gamma}$ with the density of energy of the part of the electromagnetic field coupled non minimally to gravity $\rho_{z}$ vanishes; ii) the sum of the pressure of the part of the electromagnetic field coupled minimally to gravity $p_{\gamma}$ with the pressure of the part of the electromagnetic field coupled non minimally to gravity $p_{z},$ vanishes, that is,

\begin{align}
\rho_{\gamma} + \zeta \rho_{z} = 0 \\[2ex]
p_{\gamma} + \zeta p_{z} = 0.
\label{8jan2}
\end{align}

which led to the equations
\begin{align}
3 \, X - \zeta \, ( \theta_{0} \, \dot{X} + \theta_{0}^{2} \, X ) = 0 \\[2ex]
X + \zeta \, ( \ddot{X} + 2 \, \theta_{0} \, \dot{X} + \theta_{0}^{2} \, X ) = 0.
\end{align}
Solving this system yields an explicit example of a state of the electromagnetic field that is acted by gravity but that does not modify the gravitational field. We thus obtain
\begin{align}
B^{2}(t) ~=~ constant \,  \exp{\bigl(-\frac{4}{3}\theta_{0}t\bigr)}
\end{align}

\section{Concluding remarks}

The combination of fields interacting with gravity in both, minimal and non minimal way allows the possibility to the existence of states such that in a curved geometry the corresponding energy can be annihilated. We have presented here a specific example for the electromagnetic field. However, this property is not restrict to this case. It can appears also in other contexts, for other types of matter source. A scalar field coupled minimally and non minimally to gravity also exhibit states that violates the action-reaction principle. We will come back to these cases elsewhere \cite{novellohartmann}. Furthermore, the presence of bifurcation points in the energy tensor under inspection is an example of physical situations where local causality could destroy the global causality of space-time. Maxwell's electromagnetic field, considered the prototype of classical field theory and therefore the most proeminent test-field of "deterministic" physical systems, could be responsible, when interacting directly with gravity, for generate non-deterministic cosmological situations even in the representation of perfect fluid. Therefore, the electromagnetic field, in the realm of General Relativity, may imply at least one information about the topology of the dynamical system describing the universe, namely, that it may be not unique.

\section{Appendix I: Mathematical Compendium}

Tensor $ T_{\mu\nu}$  is obtained from Lagrangian $ L_{m} $ by
\begin{align}
  T_{\mu\nu} = \frac{2}{\sqrt{- g}} \, \frac{\delta \sqrt{- g} \, L}{\delta g^{\mu\nu}}.
\end{align}

This tensor admits a representation under the form
\begin{align}
T_{\mu\nu}=\rho \, v_{\mu} \,v_{\nu}-p \,h_{\mu\nu}+q_{(\mu} \,v_{\nu)}+\pi_{\mu\nu} \, ,
\end{align}

where the 10 independent components are the energy density $\rho,$ the pressure  $ p,$ the heat flux $ q^{\alpha}$ and the anisotropic pressure $\pi_{\alpha \beta}$ given in the frame of a time-like normalized vector

\begin{align}
v^{\mu} v^{\nu}g_{\mu \nu}= 1
\end{align}

by

\begin{align}
\rho &=T^{\alpha\beta} \, v_{\alpha} \,v_{\beta} \\
p &=-\frac{1}{3} \,h_{\alpha\beta} \,T^{\alpha\beta} \\
q^{\alpha } &=h^{\alpha\beta} \, v^{\gamma}\, T_{\beta\gamma} \\
\pi^{\alpha\beta} &= h^{\alpha\mu} \, h^{\beta\nu} \, T_{\mu\nu}+p \,h^{\alpha\beta}.
\end{align}

The tensors $q_{\mu}$  and $\pi_{\mu \nu}$ satisfy the constraints
\begin{align}
q_{\mu}v^{\mu} &= 0 ,\\
\pi_{\mu \nu}v^{\mu} &= 0 , \\
\pi_{\mu \nu}g^{\mu \nu} &= 0 , \\
\pi_{\mu \nu} &= \pi_{\nu \mu}.
\end{align}

\textbf{Decomposition of any anti-symmetric tensor like Faraday tensor}
\vspace{1,25cm}

Consider an observer endowed with normalized 4-velocity $ v^{\mu}.$ He decomposes  $F_{\mu\nu}$ into electric and magnetic parts under the form:

\begin{align}
F_{\mu \nu}=-v_{\mu} \, E_{\nu}+v_{\nu} \, E_{\mu}+ \eta_{\mu \nu\rho\sigma} \, v^{\rho}\, B^{\sigma},
\end{align}

where vectors electric $(E_{\mu})$ and magnetic $(B_{\mu})$ are given by

\begin{align}
E_{\mu} &= F_{\mu \alpha} \, v^{\alpha}, \\
B_{\mu} &= F^{*}_{\mu \alpha}v^{\alpha}= \frac{1}{2} \eta_{\mu
\alpha\rho \sigma} \, F^{\rho \sigma} \, v^{\alpha}.
\end{align}

with

\begin{align}
\eta^{\alpha \beta \mu \nu}=-\frac{1}{\sqrt{-g}} \varepsilon^{\alpha \beta \mu \nu}.
\end{align}

It then follows that these vectors are orthogonal to $ v^{\mu}$

\begin{align}
E_{\mu} v^{\mu} = B_{\mu} v^{\mu} = 0.
\end{align}

The six degrees of freedom of $F_{\mu \nu}$ become represented by $(3 + 3)$ quantities associated to vectors $E_{\mu}$ and $B_{\mu}.$ We can construct two invariants

\begin{align}
 F &:= F_{\mu\nu} \, F ^{\mu\nu} \\
 G &:= F_{\mu\nu}^{*}\, F ^{\mu\nu}.
\end{align}

The algebraic identities are satisfied:

\begin{align}
^{\ast}F^{\mu\alpha}\,{}^{\ast}F_{\alpha\nu}-F^{\mu\alpha}F_{\alpha\nu}=\frac{1}{2} F \delta^{\mu}_{\ ~\nu}, \\[1ex]
\stackrel{\ast}{F^{\mu\alpha}}F_{\alpha\nu}=-\frac{1}{4} G \delta^{\mu}_{\ ~\nu}.
\end{align}

\section{Appendix II: The Weyl tensor}

Consider the general dual rotation $\mathscr{R}(\alpha)$ of the electromagnetic field (section II) and the similar one $\mathscr{R}(\theta)$ for the Weyl conformal tensor \cite{debever}

\begin{align}
\begin{split}
	\widetilde{W}_{\alpha\beta\mu\nu} &=  cos\theta \, W_{\alpha\beta\mu\nu} + sin\theta \, W^{*}_{\alpha\beta\mu\nu} \\[1ex]
\widetilde{W}_{\alpha\beta\mu\nu}^{*} &=  - sin\theta \, W_{\alpha\beta\mu\nu} +cos\theta \, W_{\alpha\beta\mu\nu}^{*}
\end{split}
\end{align}

A direct calculation shows that the quantity
\begin{align}
 L = W_{\alpha\beta\gamma\delta} \, F^{\alpha\beta} \, F^{\gamma\delta}
\end{align}
is invariant under the coupled dual rotation if the rotating angles $ \alpha $ and $\theta $ satisfy the condition
\begin{align}
 \theta + 2 \, \alpha = 0.
\end{align}

\subsection*{Acknowledgements}

We would like to acknowledge the financial support from brazilian agencies Finep, Capes, Faperj and CNPq.

\end{document}